\documentclass[twocolumn,aps,prl,groupedaddress,showpacs]{revtex4}

\usepackage{graphicx}

\begin{document}

\title{Modeling Cascading Failures in the North American Power Grid}

\author{Ryan Kinney$^{1}$, Paolo Crucitti$^2$, R\'eka Albert$^{3}$ and 
Vito Latora$^4$}
\affiliation{1. Department of Physics, University of Missouri-Rolla, 
MO 65409}
\affiliation{2. Scuola Superiore di Catania, 95123 Catania, Italy}
\affiliation{3. Department of Physics and Huck Institutes of Life 
Sciences, Pennsylvania State University, University Park, PA 16802}
\affiliation{4. Dipartimento di Fisica ed Astronomia, Universita' di
Catania and INFN, 95124 Catania, Italy}

\begin{abstract}
The North American power grid is one of the most complex technological
networks, and its interconnectivity allows both for long-distance power
transmission and for the propagation of disturbances. We model the power
grid using its actual topology and plausible assumptions about the load
and overload of transmission substations. Our results indicate that the
loss of a single substation can lead to a $25\%$ loss of
transmission efficiency by triggering an overload cascade in the network. 
We systematically study the damage inflicted by the loss of single nodes, 
and find three universal behaviors, suggesting that $40\%$ of the transmission 
substations lead to cascading failures when disrupted. While the loss of a single
node can inflict substantial damage, subsequent removals have only incremental effects, 
in agreement with the topological resilience to less than $1\%$ node loss.  
\end{abstract}

\maketitle

As demonstrated in August of $2003$, local disruptions of power
distribution within the North American power grid can result in
the loss of service to tens of millions of customers. At the heart
of this vulnerability is the fact that the power grid has developed into
one of the most complex and interconnected systems of our time, and
the same capabilities that allow power to be transferred over hundreds of 
miles also enable the propagation of local failures into grid-wide
events\cite{roadmap,b03}. To better
understand such events, the power grid needs to be studied from a
network perspective taking advantage of the recent advances in
complex network theory\cite{ab02}.

Recently a great deal of attention has been devoted
to the analysis of error and attack resilience of both artificially 
generated topologies and real world networks. The first approach that 
has been followed by researchers is that of static failures\cite{ajb00,
clmr03,aan04,mnl02,cebh00,cnsw00,cebh01} and 
consists in removing a certain percentage of elements of the systems and 
evaluating how much the performance of the network is affected by the 
simulated failure. Following such an approach it has been shown that 
the removal of a sizable group of nodes can have important consequences. 
Nevertheless, in most real transportation/communication networks, the 
breakdown of a single or of a very small size group of elements can be 
sufficient to cause the entire systems to collapse, due to the dynamics 
of redistribution of flows on the networks. To take into account this 
phenomenon, dynamical approaches have been developed \cite{hk02,ml02,mpvv03,
clm04_1,clm04_2,dcn03}. Those are 
based on the fact that the breakdown of a single component not only has 
direct consequences on the performance of the network, but also can cause 
an overload and consequently the partial or total breakdown of other components, 
thus generating a cascading effect.

Here, we use data on the network structure of the North American power
grid obtained from the POWERmap mapping system developed by Platts, the
energy information and market services unit of the McGraw-Hill 
Companies\cite{nrel}. This mapping
system contains information about every power plant, major substation,
and $115-765 kV$ power line of the North American power grid. Our
reconstructed network contains $N=14,099$ substations and $K=19,657$ transmission
(power) lines. The substations can be divided into three different groups: the generation
substations set $G_G$, whose $N_G=1,633$ elements produce electric power
to distribute, the transmission substations set $G_T$, whose
$N_T=10,287$ elements transfer power along high voltage lines, and the distribution
substations, whose $N_D=2,179$ elements distribute power to small, local
grids.As previously observed, the North American power grid forms a connected network,
thus in principle power from any generator is able to reach any distribution 
substation\cite{aan04}.

We model the power grid as a weighted\cite{wf94,lm01} graph $G$, with $N$ nodes (the substations) 
and $K$ edges (the transmission lines) and we represent it by the $N\times N$ 
adjacency matrix $\{e_{ij}\}$. The element $e_{ij}$ of such a matrix is $0$ if 
there is no direct line from the substation $i$ to the substation $j$; otherwise 
it is a number in the range $(0,1]$ that represents the efficiency of the edge.
Initially, for all existing edges, $e_{ij}$ is set equal to 1, meaning that 
all the transmission lines are working perfectly.

Both in the static and in the dynamic approach, in order to quantify how well
networks operate before and after the occurrence of  breakdowns, a measure 
of performance has to be used. Here, as in\cite{clmr03,clm04_1,clm04_2}, 
we use the average efficiency of the network\cite{lm01} that, adapted 
to the case of the North American power grid, is defined as follows: 

\begin{equation}
E=\frac{1}{N_G N_D}\sum_{i \in G_G}\sum_{ j \in G_D} \epsilon_{ij}
\label{efficiency}
\end{equation}

where $\epsilon_{ij}$ is the efficiency of the most efficient path between 
the generator $i$ and the distribution substation $j$, calculated as the harmonic
composition of the efficiencies of the component edges\cite{s88,j91,harmonic}.

Once defined the efficiency as a measure of performance, the natural definition 
of the damage $D$ that a failure causes is the normalized efficiency loss\cite{lm04}:

\begin{equation}
D=\frac{E(G_0)-E(G_f)}{E(G_0)},
\end{equation}
where $E(G_0)$ is the efficiency of the network before the occurrence of any 
breakdown and $E(G_f)$ is the final efficiency that is reached by the system 
after the end of the transient due to a breakdown, i.e. when the network 
efficiency stabilizes.

In this paper we use the dynamical approach of the CML model of Ref.\cite{clm04_1}, 
adapting it to our network.  We assume that each generator transfers
power to all the distribution substations through the transmission lines. 
The generators have also transmitting capabilities, so they are both sources
and intermediaries in power transmission. This scenario could seem 
unrealistic in the early days of electricity, when power was produced 
by local generators and transmitted only to the nearest distribution 
substations\cite{b03}. Nowadays, however, power is often redirected hundreds 
of kilometers away and our hypothesis that power from each generator 
can reach each distribution substation is not far from reality.

Adapting previous work on complex networks\cite{gkk01,n01} we define 
the load (also called betweenness) of each node
with transmitting capabilities as the number of most 
efficient paths from generators to distribution 
substations that pass through the node. As in the CML model, we associate to 
each node $i$ a capacity $C_i$ 
directly proportional to the initial load $L_i$ it carries in the unperturbed network
\cite{hk02}:

\begin{equation}
C_i = \alpha L_i(0)\quad i=1,2..N
\end{equation}

where $\alpha>1$ is the tolerance parameter that represents the ability 
of nodes to handle increased load thereby resisting perturbations.

If, due to external causes, a breakdown occurs at one or more nodes, 
so that they cannot work at all, the most efficient paths will change
and the power/load, since it cannot be destroyed, will 
redistribute among the network. Sometimes this leads to a situation 
in which a certain number of nodes, forced to carry a load higher 
than their capacity, cannot function regularly anymore  and show a 
degradation of their performance. Such a degradation can modify the 
most efficient paths, redistribute again the load on the network, 
and cause new nodes to be overloaded. If the overload caused by the 
initial breakdown is small, degradation will involve only a tiny 
part of the system, while if the overload to be reabsorbed is large 
enough, it will spread over the entire system in an avalanche mechanism, 
hindering any interaction among nodes. The degradation of performance is
represented by the following dynamical model: 

\begin{equation}
e_{ij}(t+1) = \left\{\begin{array}{rcl}
e_{ij}(0)/\frac{L_i(t)}{C_i} & \mbox {if} &  L_i(t)>C_i\\
e_{ij}(0) & \mbox {if} & L_i(t)\leq C_i\\
\end{array}\right.
\end{equation}

where $j$ extends to all the first neighbors of $i$. In other words, when a node $i$ 
is congested,  it is assumed that the efficiency of power transportation from(to) $i$ 
to(from) its first neighbors decreases linearly with the overload $L_i(t)/C_i$.

\begin{figure}[htb]
\includegraphics[width=8cm]{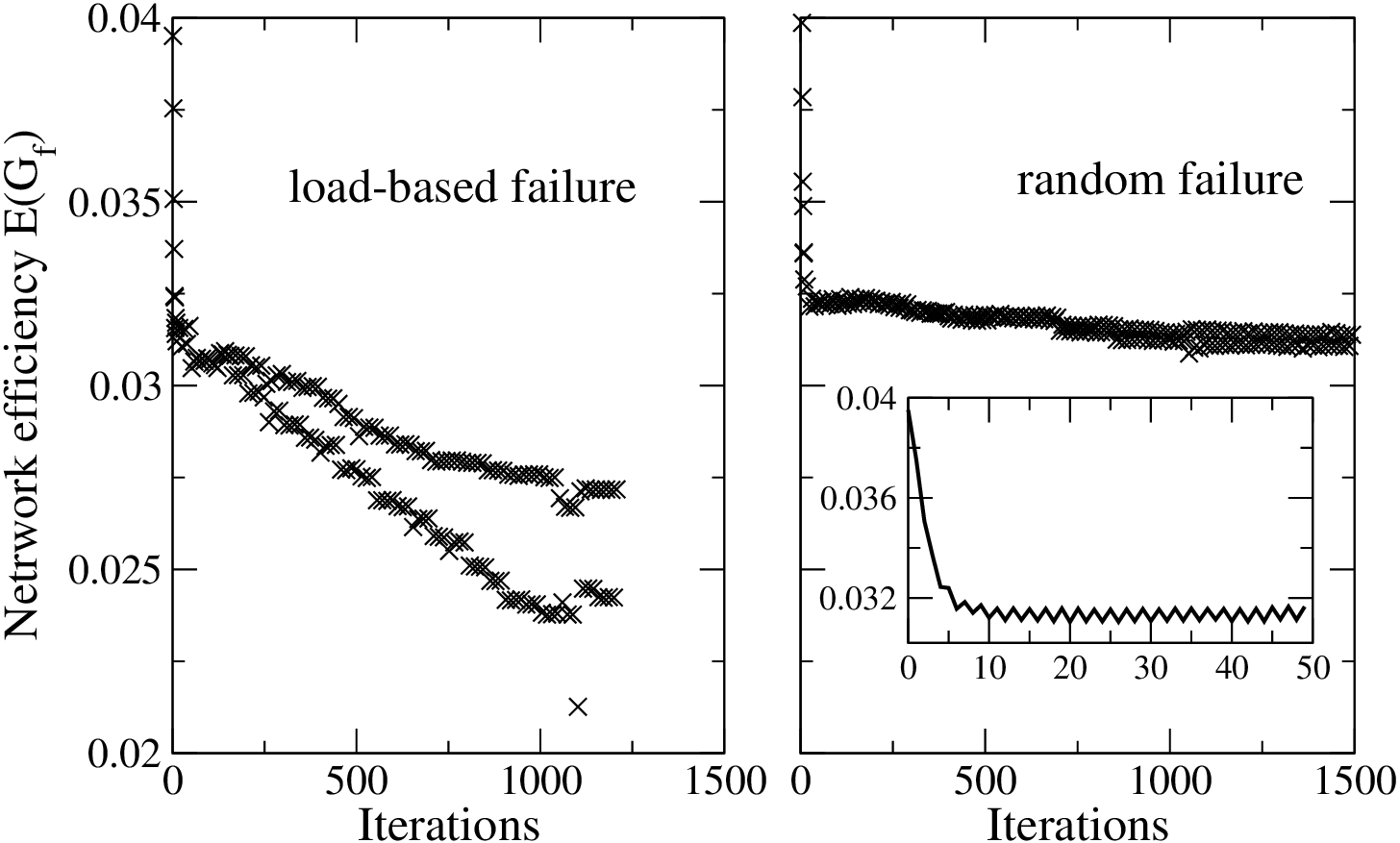}
\caption{Cascading failure with $30$ consecutive node removals.
A new node was removed at multiples of $50$ iterations, the selection was
either based on the highest load (left) or random (right). Inset: typical 
evolution of network efficiency after the removal of a single node.}
\label{cas_trans}
\end{figure}

A benefit of the CLM model is that it does not assume that overloaded nodes fail 
irreversibly, but they have the possibility of working again if, by power rerouting, 
their load decreases below their capacity. In other words the effects of overload on
nodes are reversible. Moreover, no explicit assumptions are made about
the redistribution of loads, but this redistribution emerges naturally
from the  reorganization of efficient transmission paths following a node failure. 
Simulating a network failure involves removing a node from the network and 
monitoring the progression of overloading nodes. If the tolerance parameter $\alpha$ is 
high enough the network does not present the cascading effect typical of the redistribution 
of flows and its efficiency remains unaffected by the failure. If the tolerance parameter 
is very small, instead, the cascading effect takes place and the network rapidly collapses. 
For intermediate values of $\alpha$ the network degrades more slowly and its efficiency stabilizes
to a value that is lower than the initial one. We observed that the 
efficiency of the network stabilizes into a steady state or small oscillations
around an efficiency value in about 10-20 steps. (see inset of Figure \ref{cas_trans}).

The reason for the occurrence of oscillations is strongly related to the 
reversibility of the effects of overload. Suppose that two paths 
exist from generator $i$ to the distribution substation $j$ (path A and path B) 
and that under the condition of perfect functioning (i.e. before the occurrence 
of any breakdown) path A is more efficient than path B. If at time $t$ 
some nodes of  path A become overloaded, it may happen that B becomes the most efficient path 
from $i$ to $j$. If this implies that most of the load passing through A is redirected 
to  B, the nodes of the former path will recover efficiency  to the detriment of 
some nodes of the latter one. Therefore the situation in which the most efficient 
path from $i$ to $j$ is A is restored and the redistribution of flows starts again 
its cycle. This switching between alternative paths causes the global efficiency
to oscillate. Of course in a real world network the behavior is more complicated because the 
described cycle is concurrent with a redistribution of flows that involves the 
whole network. However the oscillations are evident all the same.

In our study, we have adopted two different types of node overload progression 
schemes. The first is single node removal in which a single node is removed 
at time zero and the network is progressed in time. This way, we can model 
the effects of an external perturbation of a single transmission node or 
generator. Nevertheless, it could happen that several nodes fail at the 
same time or in close succession or are shut down to save the equipment. 
In fact, it often happens that blackouts occurs because generators and 
transformers are hardwired to protect themselves in response to a drastic 
change. To model such type of cascading failure, we develop a second node overload 
progression scheme involving many cycles of node selection and removal and 
network progression.

In both the schemes, adopting the removal strategy from\cite{clm04_1}, we have chosen 
nodes either randomly (random removal) or selectively by highest load 
(load based removal) and once removed, the efficiency of the network and 
the load of the nodes were continually recalculated in time. Only 
generation and transmission substations were removed using the above strategy.

\vspace{1.5cm}

\begin{figure}[htb]
\includegraphics[width=8cm]{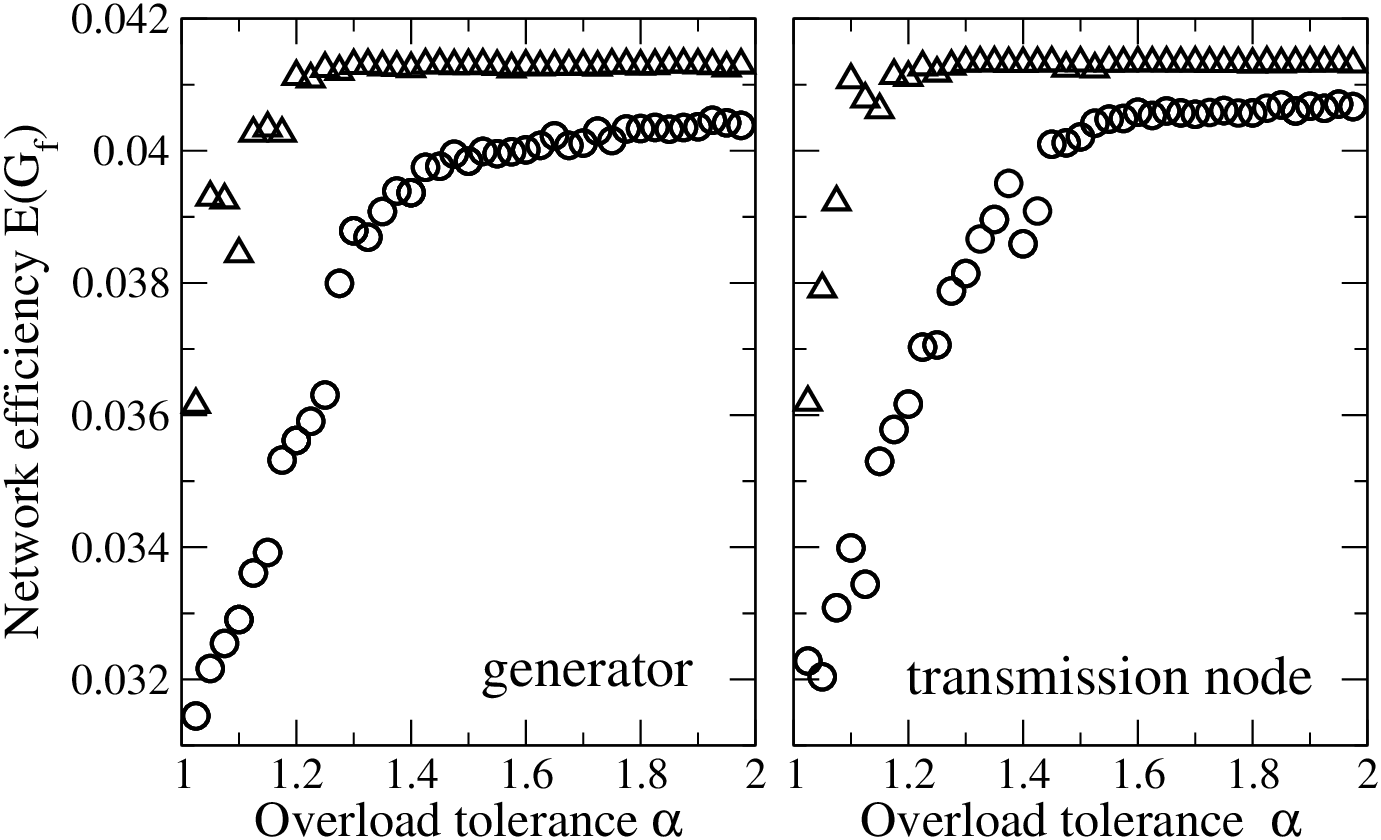}
\caption{Global efficiency of the power grid after the removal of 
random (triangles) or high-load (circles) generators or transmission substations. 
The unperturbed efficiency is $E(G_0)=0.04133$. 
As the overload tolerance $\alpha$ of the substations increases,
the final efficiency approaches the unperturbed value. The random disruption curves
were obtained by averaging over $10-100$ individual removals. The load-based disruption curve
is obtained by removing the highest load generator and transmission node, respectively.}
\label{gen_trans_com}

\end{figure}

\begin{figure}[htb]
\includegraphics[width=8cm]{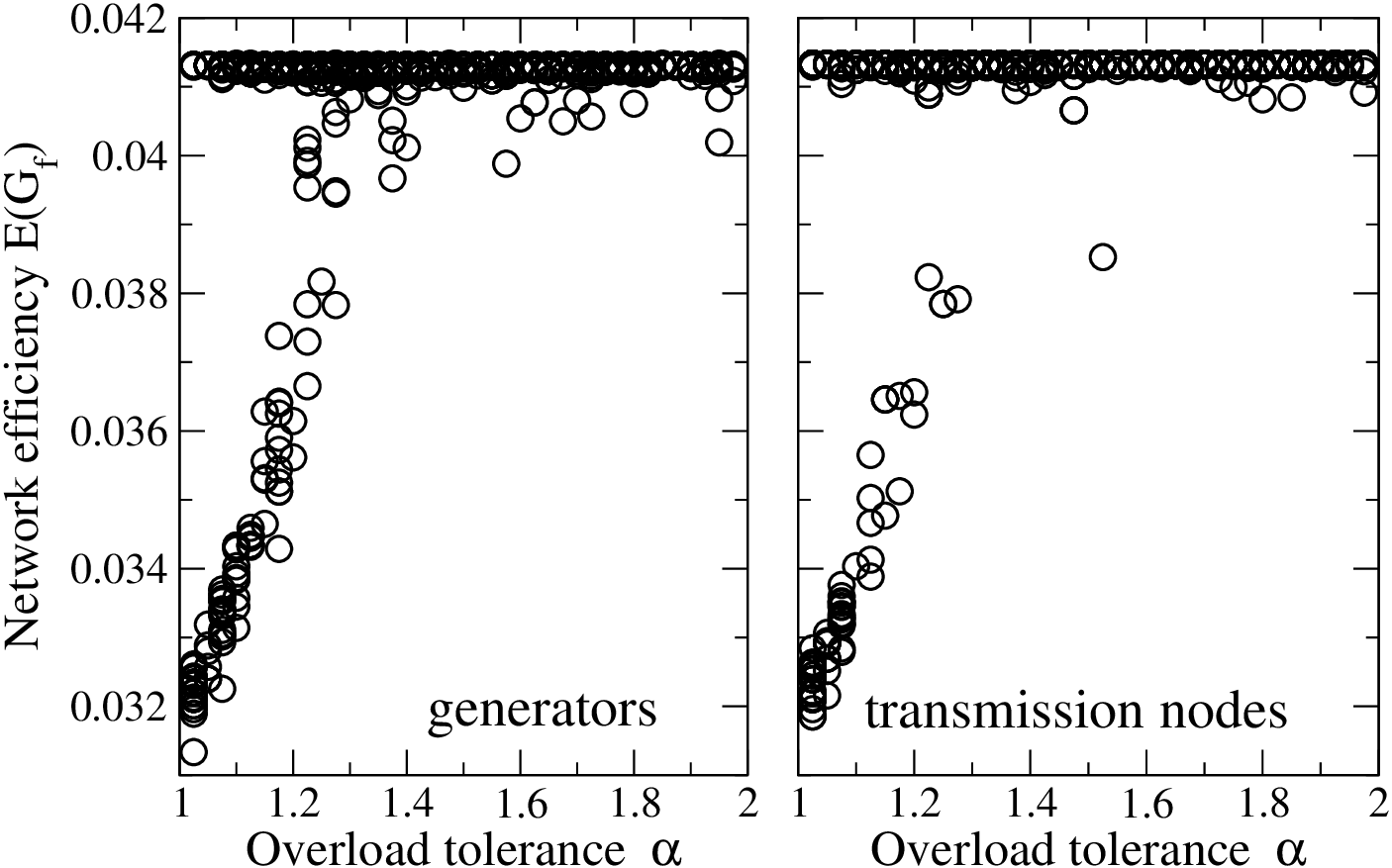}
\caption{ Scatterplot of final network efficiency for given tolerance values
for the removal of randomly selected nodes. A total of $1668$ generator and $1558$
transmission node removals are presented on this figure. }
\label{gen_trans_allran}
\end{figure}

Our first results use the single-node progression scheme for both
removal types. Figure \ref{gen_trans_com} shows a
load based (circle) removal and an average of at least $10$ random
removals (triangle) for transmission and generation substations
with final global efficiency as a function of the tolerance of the
network. These figures indicate that above a critical tolerance
value of approximately $1.42$,  the removal of the highest loaded
transmitter and generator substation has little effect on the overall 
network efficiency. However at values of tolerance below the
critical value, the global efficiency can be reduced by over
$20$ percent. For random removals, the critical value is near
$1.18$ in both figures. These results clearly indicate that the
loss of nodes with high load causes a higher damage
in the system than the loss of random nodes. Moving beyond
averages, Figures \ref{gen_trans_allran} present 
scatterplots of the efficiency of the network after the loss of
randomly selected nodes for $40$ different tolerance values. 
Two distinct trends are suggested from the efficiency versus
tolerance scatterplot. The first, a horizontal line of points
close to the unperturbed efficiency, indicates no efficiency loss
for any tolerance level. The second, corresponding to
tolerance-dependent damage, is a curve that initially increases
linearly, then saturates at high tolerance levels.

\vspace{1.5cm}

\begin{figure}[htb]
\includegraphics[width=8cm]{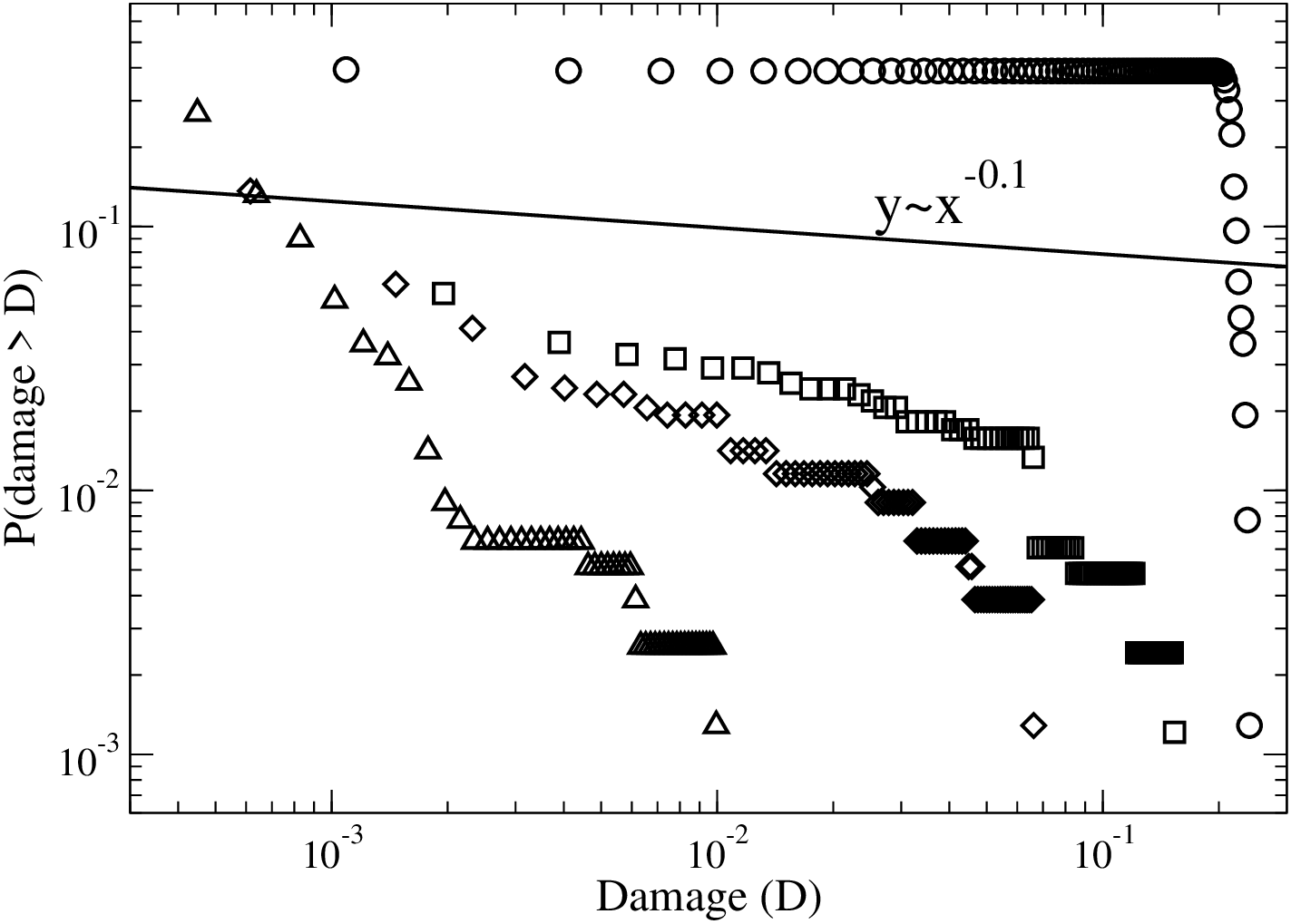}
\caption{Cumulative damage distribution for 
four different tolerance values, 
$\alpha=1.025$ (circles), $\alpha=1.2$ (squares), $\alpha=1.4$ (diamonds), 
and $\alpha=1.8$ (triangles). The continuous line indicates the cumulative
distribution of disturbances on the power grid, i.e. $P(d>D)=D^{\delta-1}$, with
$\delta\simeq 1.1$\cite{cndp00,clnd03}. }
\label{damage} 
\end{figure}

The scatterplot cannot illustrate the multiplicity of the observed
(tolerance, efficiency) points. To gain insights into the
distribution of efficiency loss we determine the cumulative damage 
distribution $P(D)$, i.e. the probability of observing damage larger than a 
given value $D$.    
Figure \ref{damage} shows the cumulative damage distribution for four
tolerance values: $\alpha=1.025$ (circles), $\alpha=1.2$ (squares), 
$\alpha=1.4$ (diamonds) and $\alpha=1.8$ (triangles). As expected, the curves corresponding to
distinct tolerance values have markedly different ranges,
indicating that the higher the tolerance value the lower the
probability to cause high damage. The long horizontal region of
the $\alpha=1.025$ curve indicates a gap between high- and low damage,
corresponding to the separation into two distinct damage behaviors
observed in the scatterplot. However, the other distributions
are relatively continuous, and all have power-law scaling regions
with exponents whose magnitude increases with tolerance, varying between $0.5$
and $2$. The probability distribution of disturbances on the power grid
has been found to be a power law with exponent close to $-1.1$\cite{cndp00,clnd03}, 
corresponding to an almost flat cumulative distribution. This is in closest agreement
with our cumulative damage distribution for low tolerances, suggesting that the overload tolerance of 
the North American power grid is low.

\vspace{1cm}

\begin{figure}[htb]
\includegraphics[width=8cm]{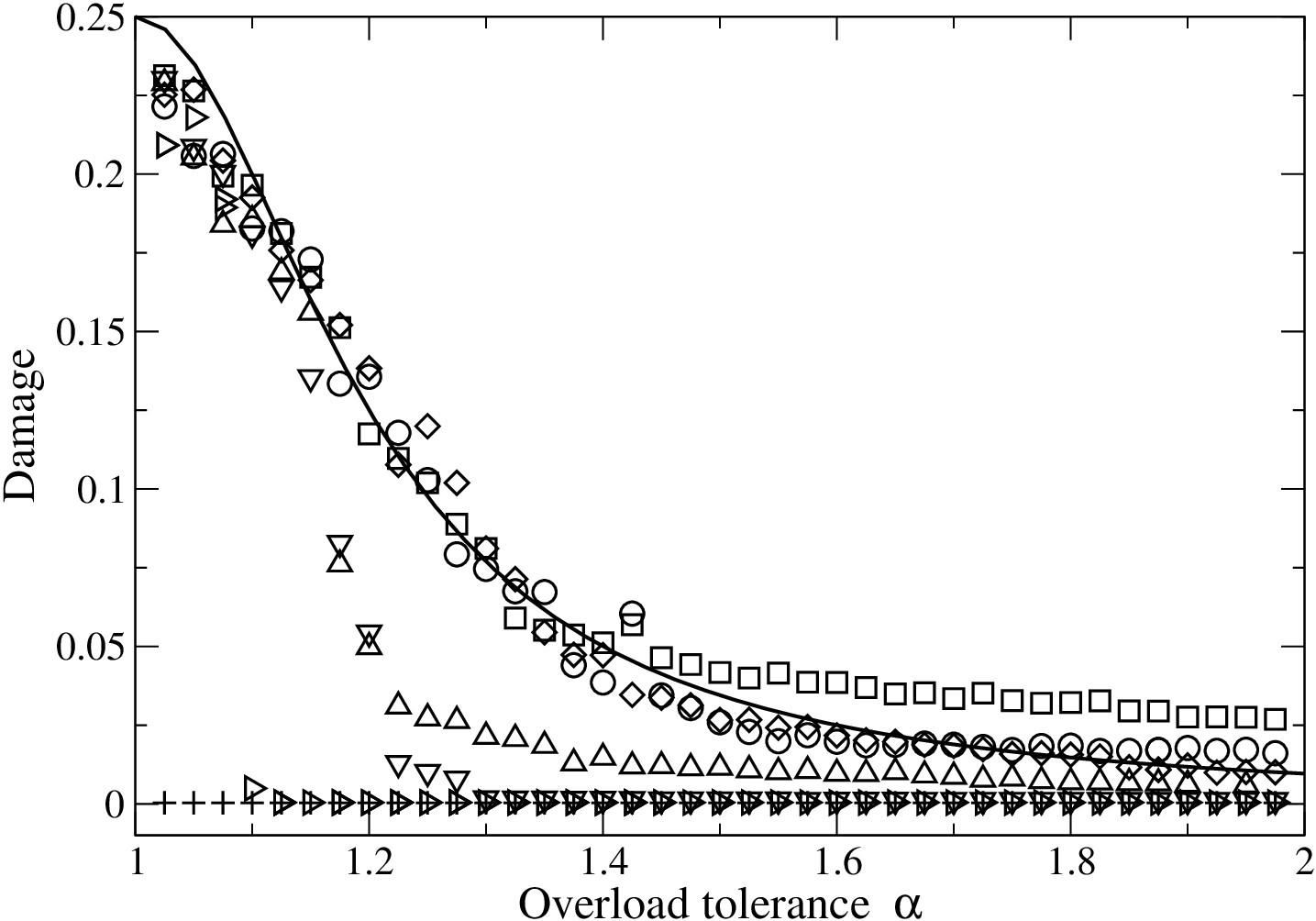}
\caption{Representative sample of node-dependent damage for different
tolerance values. Two main types of behavior can be distinguished, one corresponding
to no damage, and the other to a universal damage-versus-tolerance curve. A third type
represents a discontinuous jump from tolerance-dependent to no-damage behavior. The
continuous curve corresponds to Eq. \ref{damage_eq}.}
\label{nodeselect}
\end{figure}

Comparing  Figures \ref{gen_trans_allran} and
\ref{damage} suggests the following question: do the two distinct
(tolerance-dependent and independent) behaviors correspond to
different classes of nodes? And if the answer is yes, what
distinguishes the nodes in the two domains? To answer these
questions we selected a sample of $15$ nodes whose degrees and
loads cover the entire range of degrees and loads, and studied the
effect of their (separate) removal for a range of tolerance
values. As Figure \ref{nodeselect} shows, we find  that some
nodes' removal causes no decrease in network efficiency for the
entire range of tolerance values. Therefore, the North American power grid is
resilient to the loss of these nodes.

Included within the set of selected nodes is the node with the
highest initial load. Interestingly, the removal of that
particular node does not have the greatest effect upon the
network. The node that has the greatest effect initially and a
substantial effect over the entire range of tolerance values has
roughly $80\%$ the maximum load.

\vspace{1cm}

Based on Figure \ref{nodeselect} we conclude that there are three
separable classes of nodes:
\begin{enumerate}

\item Nodes whose removal causes no or
very little damage for any tolerance. Around $60$\% of the nodes
are in this category.

\item Nodes whose removal causes a
tolerance-dependent damage following the universal curve

\begin{equation}
D=D_{0}\left(1-\frac{x^{\beta}}{K^{\beta}+x^{\beta}}\right) 
\label{damage_eq}
\end{equation} 
where $x=\alpha -1$,
$D_{0}=0.25$ is the maximum damage, $K+1\sim 1.2$ corresponds to
the tolerance value causing half-maximal damage, and the exponent
$\beta\simeq 2$.

\item Nodes that follow the tolerance-dependent curve
for a while then suddenly jump to the no damage behavior.
\end{enumerate}

The range of damage available at a given tolerance value is from zero
(behavior 1) to the value given by the formula for behavior 2, in
good agreement with the maximum damage indicated by Figure
\ref{damage}.

\vspace{2cm}
\begin{figure}[htb]
\includegraphics[width=8cm]{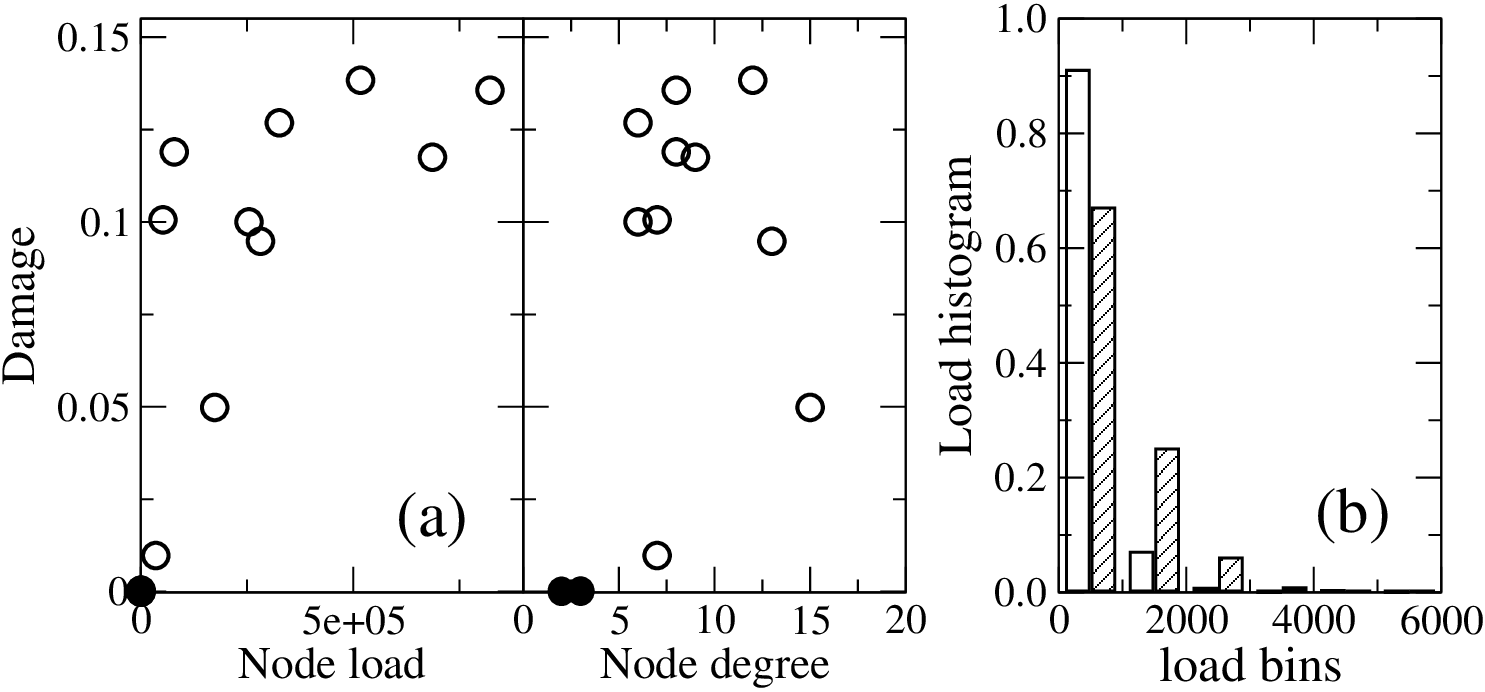}
\caption{(a) Correlation between node degree, load, and the
efficiency loss its removal causes for 15 randomly selected nodes.
The overload tolerance is $\alpha = 1.2$. The nodes causing no damage (filled circles)
have low loads and degrees.(b) Load histogram for the 
generators (white bars) and transmission substations (dashed bars) whose removal
does not cause any damage at $\alpha=1.025$.  A total
of $639$ generators and $476$ transmission substations
were included in this plot.}
\label{load_deg_eff}
\end{figure}

We find that the nodes causing no efficiency loss (behavior 1) have both
low betweenness and low degrees while the nodes that do affect the
network upon removal have higher betweenness/degree. Figure \ref{load_deg_eff}(a)
relates node degree and load with the damage caused by the node's removal
for a set of $15$ nodes. The plot indicates that, although there is no
direct correlation between degree, load and efficiency loss, nodes that have both
low degree and relatively low load will cause relatively little damage when perturbed.
Figure \ref{load_deg_eff}(b) shows the load histogram of generators and transmission
substations whose removal at tolerance $\alpha=1.025$ leads to no efficiency loss.
Each bin corresponds to a load range of $1000$. 
We find that $90\%$ of no-damage-causing generators have
loads$<1000$ and degree$<3$, while $90\%$ of non-damage-causing
transmission nodes have load$<2000$ and degree=2. The fraction of
generators with degree 1, expected to cause insignificant
efficiency loss, is $72\%$. Thus the network's resilience is
higher than trivially expected.

Moving to the cascading failure, figure \ref{cas_trans} shows
a transmitter substation load based failure at a tolerance of
$\alpha=1.025$. Here we remove the highest-load node, wait for the system
to stabilize, then find and remove the current highest-load node,
repeating this iteration several times.  Interestingly, the final
global efficiency after nearly thirty nodes removed is within ten
percent of the final global efficiency for a single-node removal
of same type and tolerance. The first node removed does the most
damage while each successive removal does little to the worsening
of efficiency. Similar behavior is recorded for generators.
In random removals most behaviors, due to the higher
probability of selecting a low degree and low betweenness node,
reach stability, where the efficiency remains roughly constant,
fter the first removal as in figure \ref{cas_trans}.

This results is complementary and similar in spirit to
the results of static transmission node removals\cite{aan04}
where the removal of up to $1\%$ of the nodes had little effect on
the connectivity of the power grid. As reference \cite{aan04} has found,
in this regime the connectivity of the grid, in other words the
reachability between generators and distribution substations, decreases approximately
proportionally with the fraction of nodes removed.
Here we obtain efficiency loss (damage) of $40\%$  after the removal
of $0.33\%$ of the high-load transmission nodes. Both of these results suggest
that perturbations higher than $1\%$ are needed for catastrophic failure.

\vspace{0.7cm}

These results are simultaneously reassuring and ominous. The North
American power grid has been proven both theoretically and
empirically to be highly robust to random failures. However, this
research highlights the possible damage done to the network by a
more targeted attack upon the few transmission substations with high betweenness and
high degree. Our results, taken together with the observed disturbance distribution,
suggest that even the loss of a single high-load and high-degree
transmission substation reduces the efficiency of the power grid by
$25\%$. This vulnerability at the transmission level deserves serious 
consideration by government and
business officials so that cost effective counter measures can be
developed. Two possibilities include reducing the load upon the
highly loaded nodes by building more transmission substations and 
controlling the spread of the cascade \cite{mo04, note}, 
or by producing power on a more local level via environmentally 
friendly methods.

\begin{acknowledgments}

 The authors wish to thank Gary L. Nakarado, Donna Heimiller and Steven 
 Englebretson for their help in obtaining the POWERmap network data. 
 The work of R. K. was supported by the Pennsylvania State University
Research Experiences for Undergraduates program. R. A. gratefully 
acknowledges a Sloan Fellowship
in Science and Engineering.

\end{acknowledgments}


\begin{thebibliography}{2}
\bibitem{nrel}
Our access to this data was made possible by the National Renewable Energy Laboratory.
\bibitem{harmonic}
The harmonic composition of $N$ numbers $x_1 ,x_2 , . . . ,x_N$  is defined as 
 $[\sum_i^N 1/x_i]^{-1}$
\bibitem{note}
Note that it is not possible to reduce the load of transmission substations by eliminating
low-load (leaf) nodes. These nodes correspond to generators with one outgoing 
power line or distribution substations that have one incoming high-voltage power
line. Consequently eliminating these nodes would diminish power generations
or disconnect it from power consumption.
\end{thebibliography}
\end{document}